\documentclass[preprint,aps,prb,superscriptaddress,citeautoscript,floatfix]{revtex4}
\usepackage{txfonts}
\usepackage{graphicx}

\newcommand{\mycite}[1]{Ref.\,\onlinecite{#1}}

\begin{document}

\title{Photon storage with sub-nanosecond readout rise time in coupled quantum wells}

\author{A.G. Winbow}
\affiliation{Department of Physics, University of California at San
Diego, La Jolla, CA 92093-0319}

\author{L.V. Butov}
\affiliation{Department of Physics, University of California at San
Diego, La Jolla, CA 92093-0319}

\author{A.C. Gossard}
\affiliation{Materials Department, University of California at Santa
Barbara, Santa Barbara, California 93106-5050}

\begin{abstract}

\textit{\textbf{The following article has been accepted by Journal of Applied Physics. After it is published, it will be found at \texttt{http://jap.aip.org/}}}
 
\medskip
Photon storage with 250\,ps rise time of the readout optical signal
was implemented with indirect excitons in coupled quantum well
nanostructures (CQW). The storage and release of photons was
controlled by the gate voltage pulse. The transient processes in the
CQW were studied by measuring the kinetics of the exciton emission
spectra after application of the gate voltage pulse. Strong
oscillations of the exciton emission wavelength were observed in the
transient regime when the gate voltage pulse was carried over an
ordinary wire. Gating the CQW via an impedance-matched broadband
transmission line has lead to an effective elimination of these
transient oscillations and expedient switching of the exciton energy
to a required value within a short time, much shorter than the
exciton lifetime.
\end{abstract}

\date{July 11, 2008}
\maketitle

Photon storage is an essential part of optical signal processing in
optical networks. Efficient photon storage in semiconductor
nanostructures has been recently demonstrated. Photons were stored
in the form of separated electrons and holes in acoustically
\cite{Rocke1997,Santos1999} and electrostatically
\cite{Zimmermann1999,Zhang2000,Krauss2004} induced lateral
superlattices, quantum dot pairs
\cite{Lundstrom1999,Kroutvar2003,Kroutvar2004,Young2007,Krenner2008},
and coupled quantum wells (CQW)\cite{Winbow2007}. The fastest
demonstrated rise time of the readout optical signal in these
devices was about one nanosecond\cite{Young2007,Winbow2007}. Here,
we report on refining the photon storage in CQW and achieving a
250\,ps rise time of the readout optical signal. The refining also
has led to an effective elimination of transient oscillations after
the storage pulse.

The storage device employs spatially separated electrons and holes
in CQW, Fig. 1a. The same device was employed in the proof of
principle photon storage in CQW\cite{Winbow2007}. The storage
is presented for low temperatures where the spatially
separated electrons and holes are bound, forming indirect excitons
(for a review on indirect excitons see \mycite{Butov2004r}); however,
the operation principle of the device is the same at high
temperatures for unbound electrons and holes. The same device and
temperature as in \mycite{Winbow2007} are studied in this paper
so that the refined photon storage can be compared with the earlier
one.

The principle of the photon storage with CQW is as follows: The
emission rate of the indirect excitons (or unbound electrons and
holes) is determined by the overlap between the electron and hole
wave functions and can be controlled by the applied gate voltage,
typically within several orders of magnitude. The energy of the
indirect excitons (or unbound pair of electron and hole) is also
controlled by the applied gate voltage $V_\mathrm{g}$, which results
in the exciton energy shift $\delta E = e d F_z$,\cite{Miller1985}
where $d$ is the separation between the electron and hole layers
(close to the distance between the QW centers), $F_z =
V_\mathrm{g}/D$ is an electric field perpendicular to the QW plane,
and $D$ is the width of the intrinsic layer in the
$n^+\!-\!i\!-\!n^+$ CQW sample. The energy can be typically
controlled within several tens of meV in CQW samples. Figs.~2a-d
present the schematic of the photon storage. Photons generated by a
laser pulse (Fig.~2a) are absorbed in the CQW device. Writing is
performed by the gate voltage pulse $V_\mathrm{g}$ (Fig.~2b), which
reduces the exciton emission rate and stores the absorbed photons in
the form of indirect excitons. Emission of the indirect excitons
during storage occurs at energy lower by approximately
$edV_\mathrm{g}/D$ (for corrections due to the interaction see
\mycite{Butov2004r}) and is weak due to their long lifetimes. This
emission is shown schematically in Fig.~2d. Readout of the stored
photons is provided by termination of the gate voltage pulse, which
increases the emission rate and results in conversion of excitons
back to photons (Fig.~2c). The earlier implementation of this scheme
\cite{Winbow2007} demonstrated photon storage with microsecond
storage time and nanosecond rise time of the optical readout.

A principal limitation of the proof-of-principle experimental setup
\cite{Winbow2007} was that the gate voltage pulse from the pulse
generator was delivered to the CQW sample via ordinary wires over
$\sim$1\,m from the top of the He cryostat to the sample at the
bottom. Such a circuit segment is suited for dc signals. However,
for the nanosecond switching time of the applied gate voltage, i.e.
GHz switching speed, the characteristic wavelength $\sim{}c / f$ is
smaller than the $\sim$1\,m length of the wire. An impedance-matched
broadband transmission line is required in such regime for optimal
device performance. This is briefly discussed below.

The characteristic impedance of the pulse generator and external
cabling is $Z=50\,\Omega$, while the CQW sample has resistance
between the top and bottom planes in the range of M$\Omega$ -
G$\Omega$ depending on the laser excitation and acts approximately
as a parallel plate capacitor with $C \sim \epsilon S/4\pi D \sim$
60\,pF, where $S \sim 0.5 \times 0.5$\,mm$^2$ is the sample area.
(Note that such circuit also acts as an $RC$ low-pass filter, which
slows the switching time.) The apparent impedance mismatch between
the ordinary wire and the sample results in reflection of the
voltage pulse at the sample and oscillation of the applied voltage
$V_\mathrm{g}$ and electric field $F_z=V_\mathrm{g}/D$ in the
sample. We detected these oscillations by analysis of the evolution
of the emission spectra, as described below. (The voltage
oscillations were also observed on an oscilloscope.) These
oscillations hinder the storage in two ways: First, each swing of
the oscillating electric field reduces the exciton lifetime and,
therefore, reduces the photon storage efficiency. Second, the
oscillations complicate the readout process, while waiting for their
damping (as in the proof-of-principle experiment \cite{Winbow2007})
sets a minimum storage time.

Furthermore, an ordinary wire also acts as an antenna. In the case
when multiple gate voltages are applied to the CQW sample via
different wires, the radiation emitted by such an antenna can lead
to crosstalk among the wires. Multiple wires are used for creating
potential landscapes for excitons, as discussed below, and thus
eliminating crosstalk is required for improving the control of
potential landscapes for excitons.

Therefore, improving the device performance requires gating the CQW
via an impedance-matched broadband transmission line appropriate for
the demanded switching speed. In this paper, we exploit gating the
CQW sample via a broadband coaxial cable with a 50\,$\Omega$
termination resistor, see Fig.~1b. This method can be applied to a variety of semiconductor structures of diverse layer designs. The achieved performance improvement of our device is described below.

$n^+\!-\!i\!-\!n^+$ GaAs/AlGaAs CQW samples were grown by molecular
beam epitaxy. The $i$ region consists of a single pair of 8\,nm GaAs
QWs separated by a 4\,nm Al$_{0.33}$Ga$_{0.67}$As barrier and
surrounded by 200\,nm Al$_{0.33}$Ga$_{0.67}$As barrier layers. The
$n^+$ layers are Si-doped GaAs with $N_{\mathrm{Si}}=5 \times
10^{17}$\,cm$^{-3}$. The electric field in the sample growth
direction $F_z$ is controlled by the gate voltage $V_\mathrm{g}$
applied between $n^+$ layers. At $V_\mathrm{g}=0$, the lowest energy
state in the CQW is the direct exciton with a short lifetime, while
at $V_\mathrm{g} \sim 1.4$\,V, the lowest energy state is the
indirect exciton with a long lifetime, about 0.1\,$\mu$s for the
studied sample.

The carriers were photoexcited by a 635\,nm laser diode. The 200\,ns
laser excitation pulse (Fig.~2a) has a rectangular shape with edge
sharpness  $\sim 0.6$\,ns and repetition frequency 500\,kHz. The
average excitation power was $80\,\mu$W and the excitation spot
diameter was $\sim 100\,\mu$m. The emitted light was diffracted by a
single-grating spectrometer and detected by a Peltier-cooled
photomultiplier tube and time correlated photon counting system. The
experiments were performed in a He cryostat at $T\approx 5$\,K.

The step-like voltage pulses for the storage and read-out were
provided by a pulse generator with 0.5\,ns exponential rise/fall
time. (Note that the rise time of the readout signal was shorter
than 0.5\,ns; after termination of the gate voltage pulse, the
emission line moves to higher energies to reach the energy at zero
electric field and the rise time of the readout optical signal
measured at this energy depends on the spectral shape of the
emission line and is faster for the lines with a sharp high-energy
edge.) The pulse was transmitted within the cryostat over a
semi-rigid coaxial cable UT-141B-SS with silver-plated
beryllium-copper inner conductor, PTFE teflon dielectric, and
stainless-steel outer shell of diameter 3.6\,mm, having
room-temperature attenuation of 3\,dB/m at 10\,GHz. The cable
bandwidth complies with the requirement for fast control while the
cable composition reduces heat conductance to the sample, thus
facilitating future measurements with several such cables at low
temperatures. The cable was routed vertically straight from the
sample through the exterior vacuum SMA feedthrough and countersunk
in a channel in the sample plane for good electrical ground. The 5\,mm 
contact pin from the sample socket passed through a hole in the
ground plane and terminated on the back side with a metal-film
surface-mount resistor having 50\,$\Omega$ at 4.2\,K.

Figure~2f shows that the rise time of the readout optical signal in
the refined system was 250\,ps, which is an improvement compared to
the nanosecond rise time achieved in earlier studies.

We also analyzed the transient processes in the CQW after the
storage pulse by measuring the kinetics of the exciton emission
spectra. Figures~3c,d show the presence of strong oscillations of
the exciton emission wavelength in the transient regime for the CQW
gated via dc-suited wiring. The oscillation of the emission energy
$E$ reveals the oscillation of the electric field in the sample
$F_z$, with the relation given by $\delta E = e d \delta F_z$. After
the voltage pulse, the emission wavelength of the indirect excitons
varies from about 810\,nm ($E\approx 1.530$\,eV) to 793\,nm ($E
\approx$ 1.563\,eV), close to the wavelength of the spatially direct
emission at 791\,nm ($E\approx$ 1.567\,eV). The emission intensity
rises strongly as the wavelength reaches a minimum, caused by the
increased recombination rate at smaller $F_z$; this causes losses of
stored photons.

Figures~3a,b show these oscillations are effectively suppressed when
the CQW is gated via an impedance-matched broadband transmission
line. After application of the gate voltage pulse, the exciton
energy changes without oscillations to the value determined by the
applied voltage. The following slow fall in energy observed in
Figs.~3a,b is consistent with the reduction of the indirect exciton
density with decay, which results in the reduction of the repulsive
interaction between the excitons \cite{Butov2004r}. The absence of
oscillations demonstrates expedient switching at storage. As
mentioned above, the oscillations cause losses of stored photons and
complicate the readout processes; their elimination improves the
device performance. Note that the data reported here demonstrate
proof of principle for refining the device performance. The issues
essential for practical applications, such as device operation at
high temperatures, are briefly discussed in \mycite{Winbow2007}.

We would like to emphasize an important application of the rapid
control of the indirect exciton energy and lifetime. A laterally
modulated gate voltage $V_\mathrm{g}(x,y)$ created by a pattern of
electrodes on a sample surface can form a variety of in-plane
potential landscapes for indirect excitons in CQW. Particular cases
for such potential landscapes include potential gradients
\cite{Hagn1995,Gartner2006}, 1D
\cite{Zimmermann1999,Krauss2004,Zhang2000,Zimmermann1997} and 2D
\cite{Hammack2005} lateral superlattices, traps
\cite{Huber1998,Hammack2005,Chen2006,High2008}, and excitonic
circuits \cite{High2007}. The switching of the exciton energy to a
required value without oscillations and within a short time, much
shorter than the exciton lifetime, Figs. 3a,b, demonstrates an
improvement for the control of such potential landscapes, which can
be exploited in studying the physics of excitons.

Note also that the storage scheme can be used to realize a cold gas
of \textit{direct} excitons.
Due to their short lifetime, direct excitons are typically
hotter than the lattice, while indirect excitons live long enough to cool down
essentially to the lattice temperature. \cite{Butov2004r} In the
storage scheme, initially hot direct excitons transform to indirect
excitons by the voltage pulse, cool down toward the lattice temperature
during the long storage time, and then transform to direct
excitons at the pulse termination. This method uses the long lifetime of indirect
excitons to realize a cold gas of direct excitons after the last step, provided that
essentially no heating occurs then. Note that indirect excitons
have a built-in dipole moment and, therefore, interact relatively
strongly \cite{Yoshioka90,Zhu95,Lozovik96,Ivanov02,Schindler2008}. However,
direct excitons have no built-in dipole moment and interact weakly.
Therefore, this method may permit extending the studies of cold
exciton gases to a new system of weakly interacting cold direct
excitons.

In conclusion, gating the CQW via an impedance-matched broadband
transmission line has lead to an effective elimination of the
transient oscillations in the electric field across the sample and
to expedient switching of the exciton energy to a required value
within a short time, much shorter than the exciton lifetime. A rise
time of the readout optical signal as short as 250\,ps was achieved.

This work is supported by ARO, DOE, and NSF. We thank K.L.~Campman
for growing the high quality samples, and R.~Heron, G.~Kassabian,
B.~Naberhuis, and R.~Parker for help in preparing the experiment.

\clearpage

\begin{figure}
\includegraphics{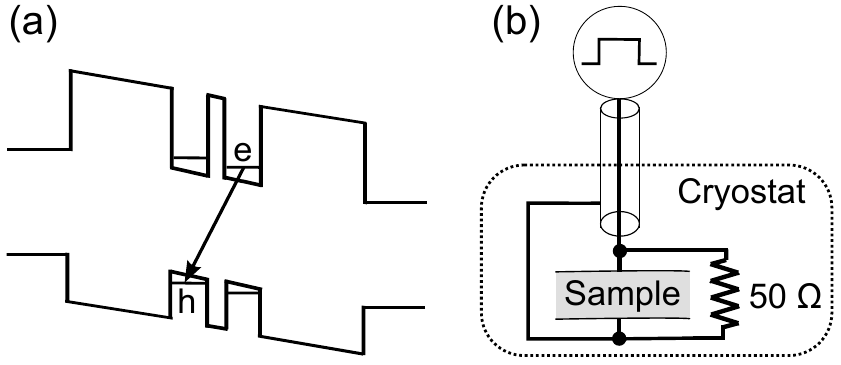}
\caption{(a) GaAs/AlGaAs CQW band diagram. (b) Schematic of the
photon storage circuit with an impedance-matched broadband
transmission line.}
\end{figure}

\begin{figure}
\includegraphics{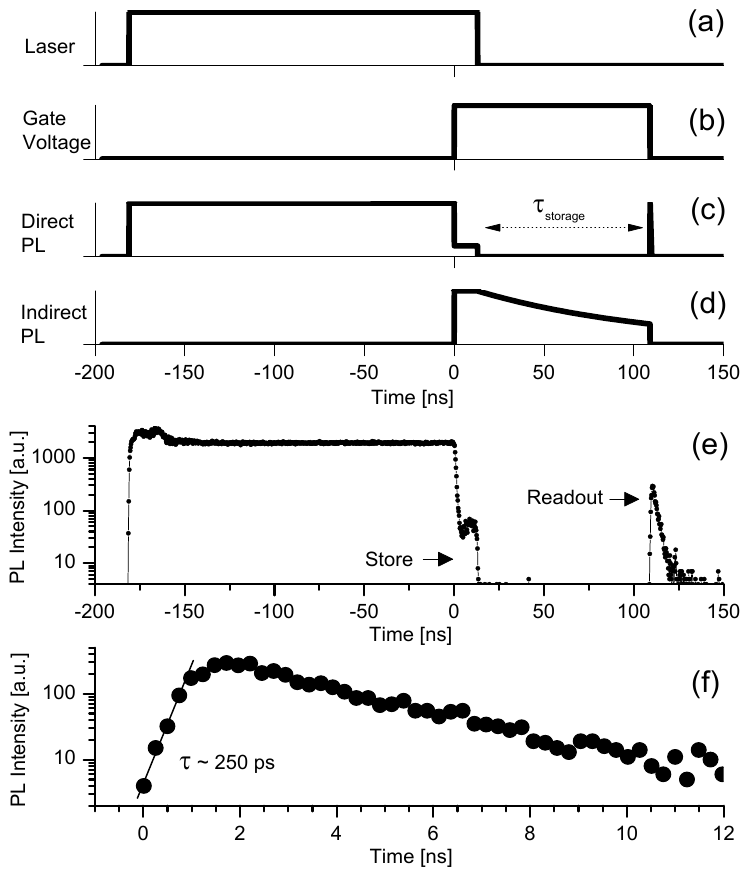}
\caption{(a)--(d) Schematic of the photon storage and readout in the
CQW device showing the sequence of the laser (a) and gate voltage
(b) pulses as well as the emission of direct (c) and indirect (d)
excitons. The operation principle of the device is described in the
text. (e,f) Experimental implementation of the photon storage in the
CQW device. The gate voltage pulse $V_\mathrm{g}=1.4$\,V. (e)
Kinetics of the direct exciton emission of the refined system with
an impedance-matched broadband transmission line, demonstrating
photon storage followed by readout. The timescale on the schematic
(a-d) corresponds to the experimental data. (f) The readout pulse on
an extended time scale. The exponential fit (line) gives the rise
time $\tau_{\mathrm{readout}} \sim 250$\,ps. The fall time and FWHM
of the readout signal are both 2.5\,ns.}
\end{figure}

\begin{figure*}
\includegraphics{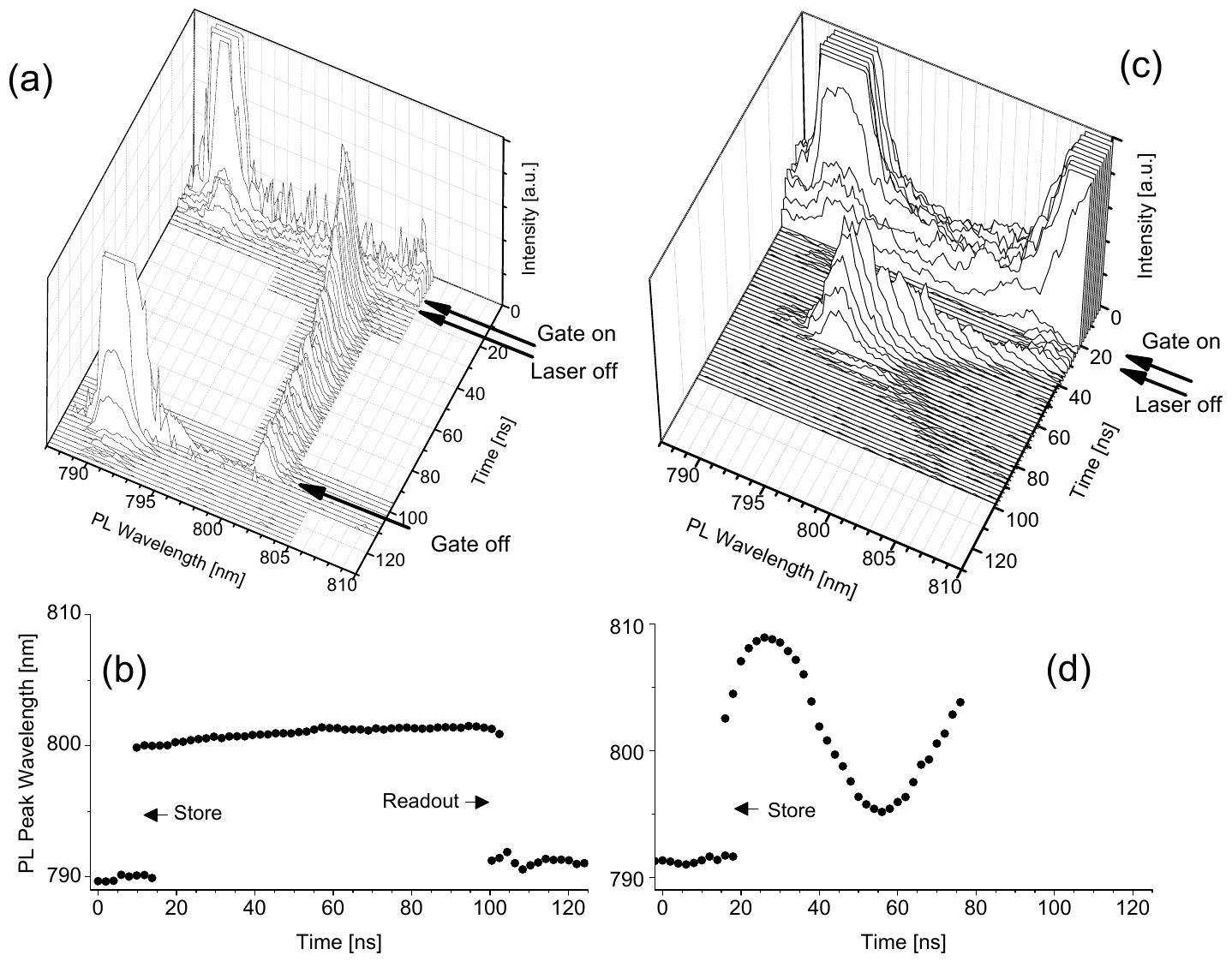}
\caption{Evolution of emission spectra during storage and readout
for (a,b) the refined system employing an impedance-matched
broadband transmission line and (c,d) the proof-of-principle system
employing dc-suited wiring. The applied voltage is
$V_\mathrm{g}=1.4$\,V. Each spectrum is stepped 2\,ns apart in time
and is measured within a 2.3\,ns window. The emission peaks while
the laser is on (a,c) and at readout (a) are clipped in the plots to
show small scale details. In the proof-of-principle system, the
readout PL pulse occurs later in time and is not shown. Bulk $n^+$
PL is seen at $\lambda \gtrsim 805$\,nm during the laser pulse in
(c). (b,d) The mean emission peak wavelength
$\bar{\lambda}_{\mathrm{peak}} = \int{\lambda
I(\lambda)\mathrm{d}\lambda} ~/ \int{I(\lambda) \mathrm{d}\lambda}$
for the spectra presented in (a,c). The refined system eliminates
the emission oscillations suffered by the proof-of-principle system
after the voltage pulse.}
\end{figure*}


\begin{thebibliography}{99}
\bibitem{term}
The term photon storage is used in a general sense and the issue of
the carrier or phase information loss is not addressed here.
\end{thebibliography}
\end{document}